\documentclass[twocolumn,superscriptaddress,showpacs,amsfonts,amsmath,amssymb,prb]{revtex4}
\usepackage{graphicx} 
\usepackage{subfig, textcomp} 
\usepackage{dcolumn} 
\usepackage{bm} 

\begin{document}

\title{Transition between 1D and 0D spin transport studied by Hanle precession}

\author{M. Wojtaszek} \email{m.wojtaszek@rug.nl} \affiliation{Physics of Nanodevices, Zernike Institute for Advanced Materials, University of Groningen, Groningen, The Netherlands} 
\author{I. J. Vera-Marun} \affiliation{Physics of Nanodevices, Zernike Institute for Advanced Materials, University of Groningen, Groningen, The Netherlands} 
\author{B. J. van Wees} \affiliation{Physics of Nanodevices, Zernike Institute for Advanced Materials, University of Groningen, Groningen, The Netherlands}

\date{\today}

\begin{abstract} 
The precession of electron spins in a perpendicular magnetic field, the so called Hanle effect, provides an unique insight into spin properties of a  non-magnetic material. In practice, the spin signal is fitted to the analytic solution of the spin Bloch equation, which accounts for diffusion, relaxation and precession effects on spin. The analytic formula, however, is derived for an infinite length of the 1D spin channel. This is usually not satisfied in the real devices. The finite size of the channel length $l_{\text{dev}}$ leads to confinement of spins and increase of spin accumulation. Moreover, reflection of spins from the channel ends leads to spin interference, altering the characteristic precession lineshape. In this work we study the influence of finite $l_{\text{dev}}$ on the Hanle lineshape and show when it can lead to a two-fold discrepancy in the extracted spin coefficients. We propose the extension of the Hanle analytic formula to include the geometrical aspects of the real device and get an excellent agreement with a finite-element model of spin precession, where this geometry is explicitly set. We also demonstrate that in the limit of a channel length shorter than the spin relaxation length $\lambda_{s}$, the spin diffusion is negligible and a 0D spin transport description, with Lorentzian precession dependence applies. We provide a universal criterion for which transport description, 0D or 1D, to apply depending on the ratio $l_{\text{dev}}/\lambda_{s}$ and the corresponding accuracy of such a choice. 

\end{abstract}


\maketitle 

\section{\label{sec:Introduction}Introduction} 
Spintronic devices  require in addition to efficient spin injection and detection components also a good spin transmitting material, with minimal loss of spin information \cite{Fabian2007}. The proper extraction of spin properties in a particular material: spin relaxation time $\tau_{s}$, spin diffusion coefficient $D_{s}$ and spin relaxation length $\lambda_{s}=\sqrt{\tau_{s}D_{s}}$, is essential for understanding of spin dephasing mechanisms and for improving the spin performance of materials. So far the most promising materials for electronic spin guide are: Silicon\cite{Suzuki2011} with $\lambda_{s}\simeq 0.5$~$\mu$m  and graphene with $\lambda_{s}\simeq 5\ \mu$m at room temperature\cite{Zomer2012}. 

Spin properties are mostly characterized by the electronic transport in a lateral spin valve structure, which is a device with ferromagnetic injector and detector separated by the nonmagnetic material of interest \cite{Johnson1988,Jedema2003, Tombros2007, Zomer2012, Suzuki2011, Idzuchi2012}. 
Hanle effect, especially in four-terminal nonlocal geometry is considered the most reliable way for determining spin coefficients \cite{Johnson1988a,Jedema2002}. Firstly, it enables the spin signal to be separated from a spurious background and, secondly, it allows $\tau_{s}$ and $D_{s}$ to be independently extracted . 
Although very powerful, the fitting of the Hanle measurements requires much care. Effects of invasive contacts on Hanle signal are discussed in Ref.~\onlinecite{Maassen2012b}. An influence of the finite device geometry has so far been discussed only in its relation to the amplitude of the spin accumulation \cite{Jedema2003, Jaffres2010} ignoring the effects on the Hanle signal. However the geometrical confinement leads to interference effects due to reflection of spins from the channel ends, which modifies the lineshape of the Hanle signal critical for extracting $\tau_{s}$ and $D_{s}$. 

Here we address the spin confinement effects, in particular the compatibility of the standard analytic Hanle formula, derived for infinite channel length, with the signal obtained in devices of finite length $l_{\text{dev}}$. This situation is common when dealing with micromechanically cleaved graphene \cite{Tombros2007, Zomer2012}. 
We determine how the finite length of the spin channel alters the Hanle precession lineshape and the spin coefficients extracted when fitting with a standard analytic solution. We establish the length scale at which the device size is important and propose an analytic extension of the Hanle formula to include the multiple reflections of spins from the channel ends. We verify our analytic model with a finite-element numerical model of spin precession, where the device geometry is explicitly defined. Our analytic model can be easily adapted to any lateral device geometry. We also indicate when the confinement of the spins results in a transition from 1D to 0D spin transport description with a simple Lorentzian-type of Hanle. Our work concludes with instructions on how to account for the finite geometry effects in the Hanle fitting procedure. We show that the role of geometry scales universally with $\lambda_{s}$. The presented description can be applied to any channel material like non-magnetic metals or semiconductors, including graphene.


\section{\label{sec:Analytic_single_Hanle}Hanle effect in an infinite channel} %
Spin transport through a non-magnetic material can be successfully modeled by distinguishing two spin channels \cite{Valet1993} - one for majority  and one for minority spins - and by introducing the spin accumulation $\mu_{s} = (\mu_{\uparrow}-\mu_{\downarrow})/2$, which is the difference between the electrochemical potential for each spin channel. We focus the analysis on non-local spin transport, where the current flows through the part of the device that is outside of the detection circuit.
We base our analysis on graphene as a channel material to reduce the spin transport description to a 2D case. Moreover, in strips, where contacts cover the whole width of the channel, the spin profile is uniform along the contact and the description can be further reduced to 1D. We also exclude conductivity mismatch effects and contact induced dephasing \cite{Takahashi2003, Maassen2012b}, as they can be circumvented by setting a highly resistive tunnel barrier.

\begin{figure}
\centering
  \subfloat[]{  
     \includegraphics[width=0.45\columnwidth]{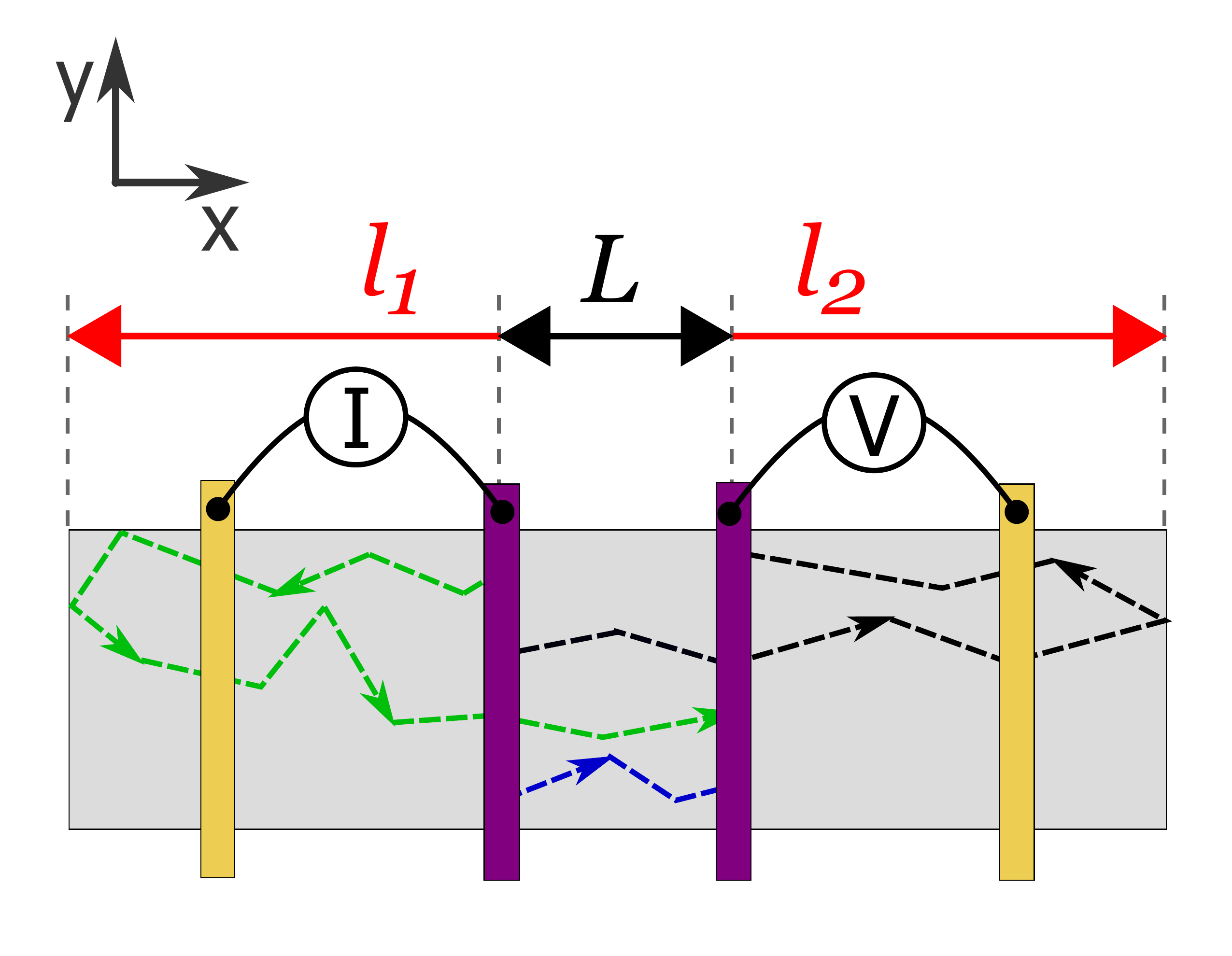} 
   \label{fig:spin_random_walk}}
  \subfloat[]{  
   \includegraphics[width=0.45\columnwidth]{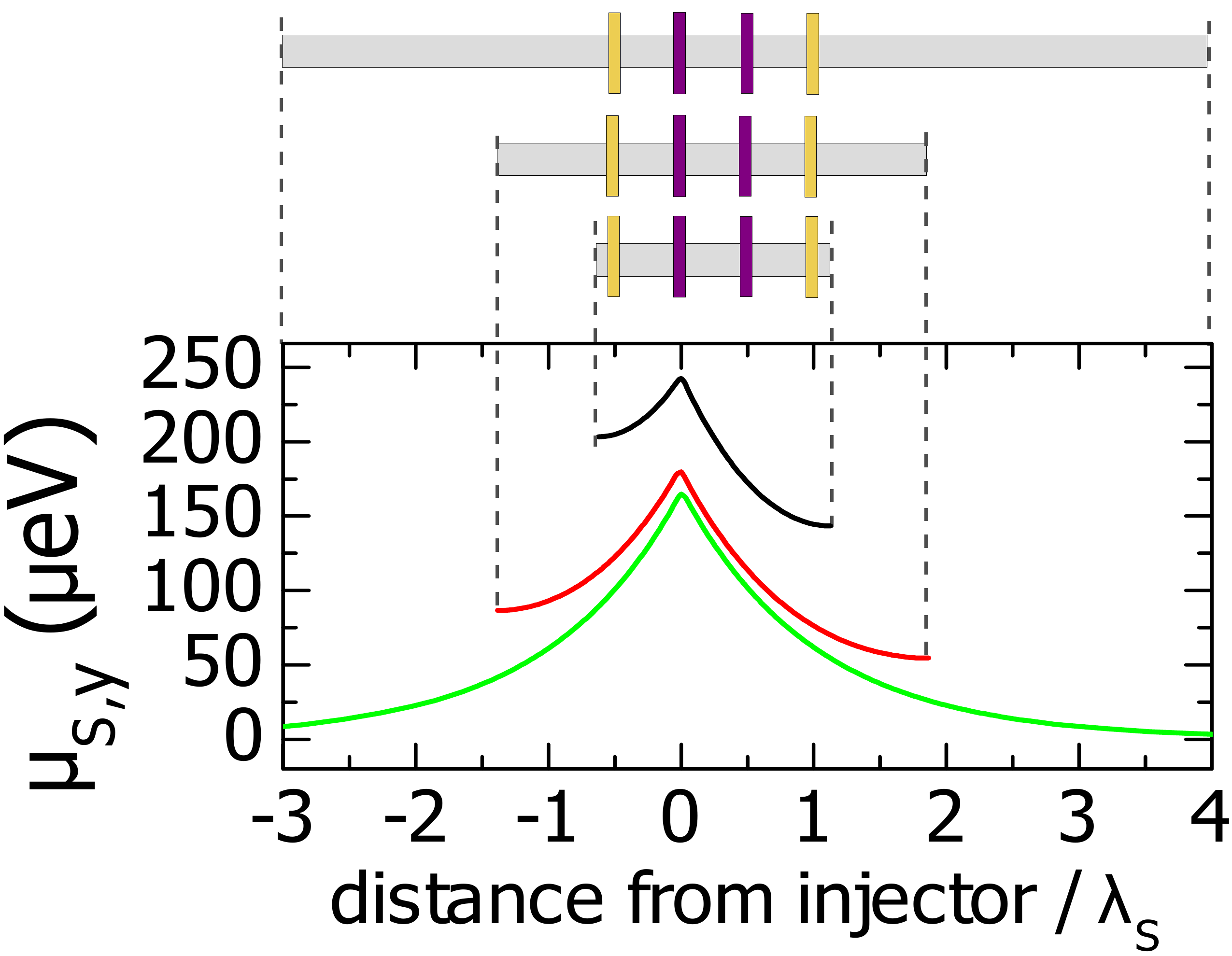} 
   \label{fig:spin_profile_vsL}}
   \\
  \subfloat[]{  
   \includegraphics[width=0.45\columnwidth]{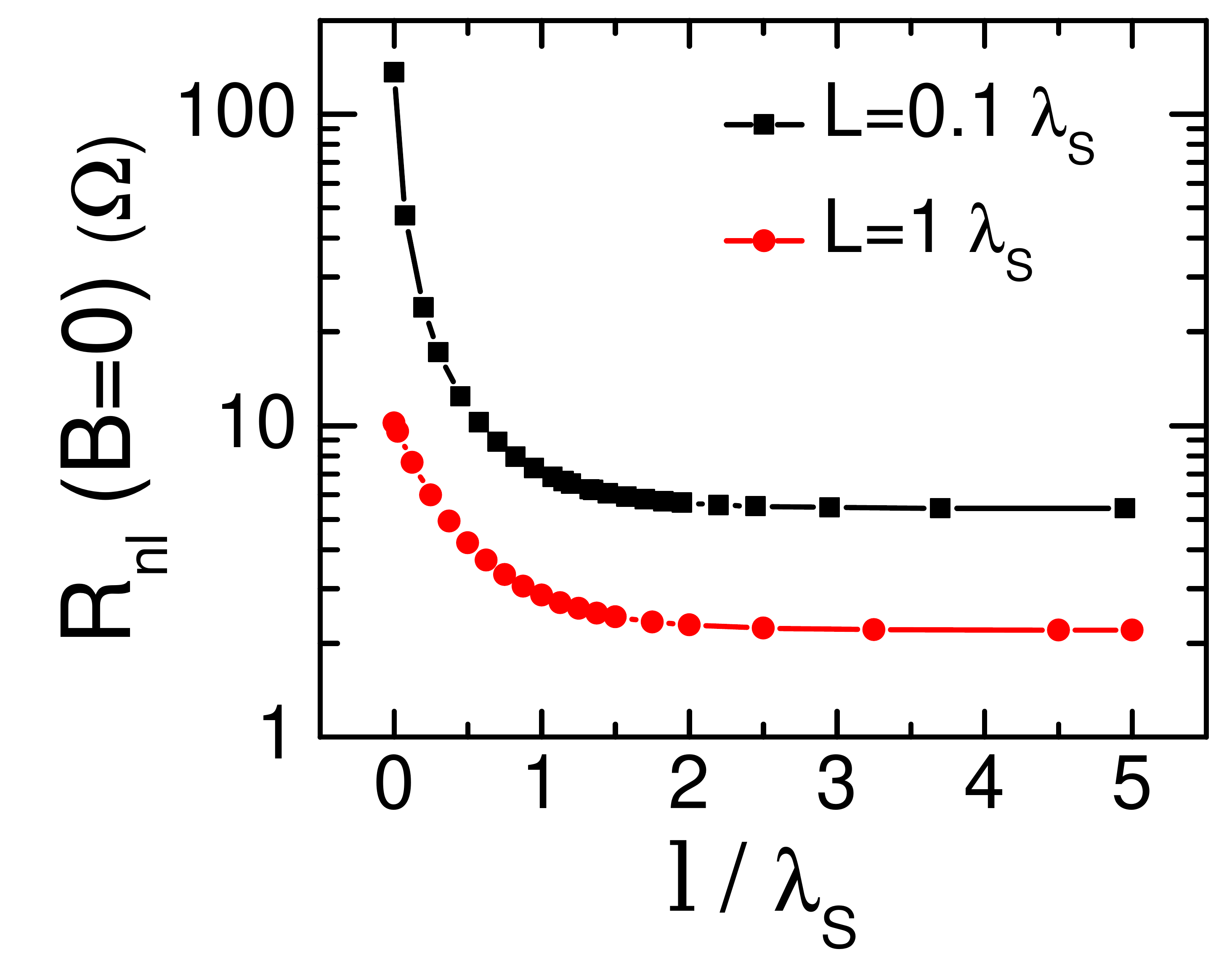} 
   \label{fig:Hanle_maximum}}
  \subfloat[]{  
      \includegraphics[width=0.45\columnwidth]{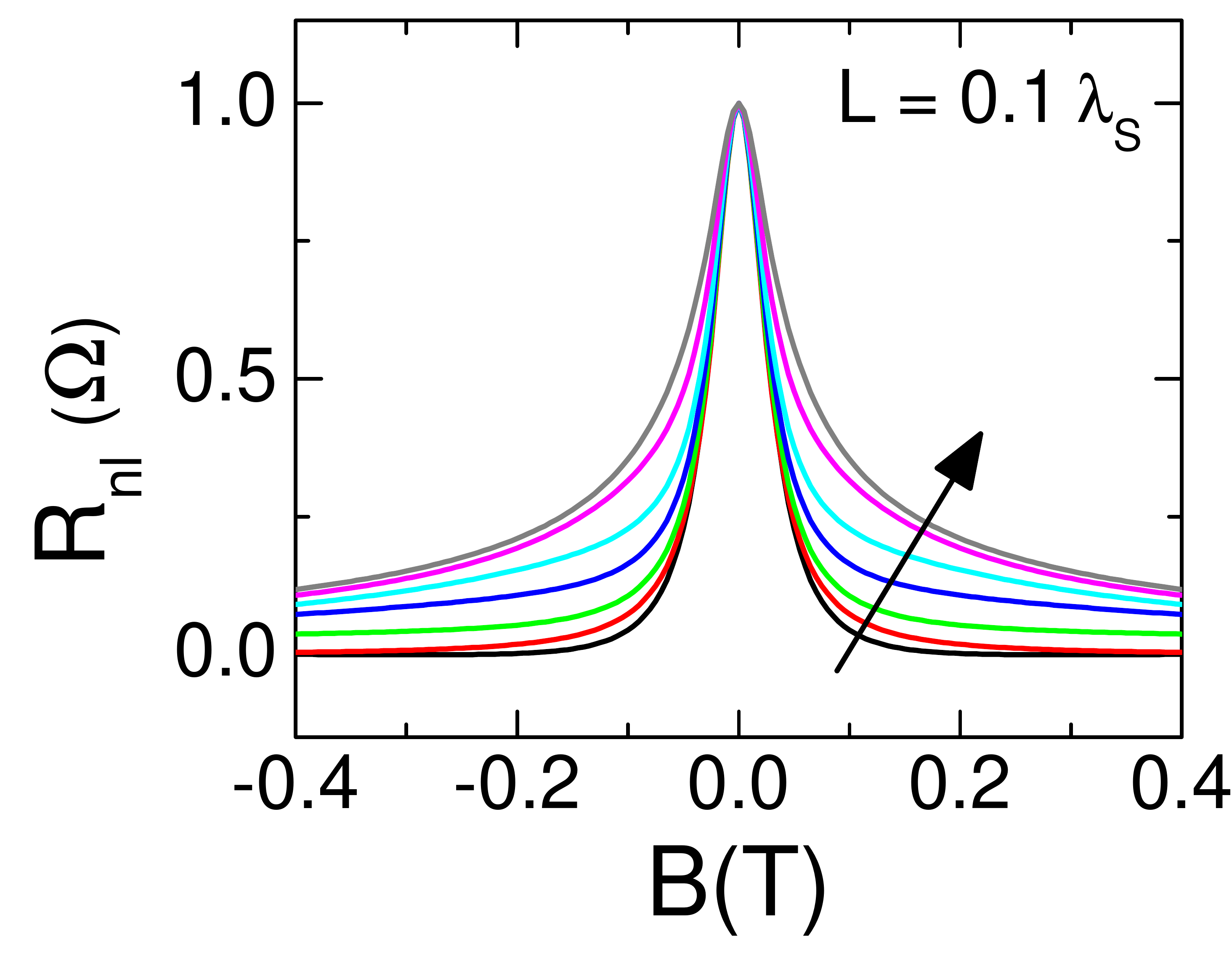} 
      \label{fig:norma_Hanle_lineshape}}
\captionsetup{format=hang,justification=centerlast, font=small}
\caption[]{(Color online) (a) Scheme of the device geometry and non-local transport measurement, with 2 ferromagnetic inner electrodes and 2 non-magnetic outer electrodes. The dashed lines indicate few possible trajectories for the spin, for which $d=L, L+2l_{1},L+2l_{2}$. All the contacts are higly resistive and do not induce spin relaxation. (b) Numerical evaluation of $\mu_{s,y}$ along the spin channel for 3 different device lengths at $B$=0. (c) Amplitude of $\Sigma R_{\text{nl}}$ from Eq.~\ref{eq:Rnl_Hanle_sum} as a function of channel extension $l_{1}=l_{2}=l$. (d) Normalized $\Sigma R_{\text{nl}}$ for different $l_{\text{dev}}$ and constant $L=0.1 \lambda_{s}$. Following the arrow direction are curves for $l_{\text{dev}}/\lambda_{s}=\{0.1, 0.5, 1, 1.5, 2, 3,  10\}$. Only the last one can be properly fitted with the standard Hanle formula from Eq.~\ref{eq:Rnl_Hanle}. 
}
\end{figure}

The device geometry and the four-terminal nonlocal measurement scheme considered here is presented in Fig.~\ref{fig:spin_random_walk}.  The outer electrodes (current drain and reference probe) are non-magnetic, therefore in a linear transport regimes their specific location does not matter \cite{Vera-Marun2011}. 
The injected spins are oriented in-plane following the magnetization direction of the ferromagnetic injector along the $y$-axis. When one applies a perpendicular magnetic field $B=B_{z}$ the spins start a precession in the $x-y$ plane. 
 
In the steady state these processes can be described by the one-dimensional Bloch equation for $\boldsymbol{\mu_{s}}=(\mu_{s,x}, \mu_{s,y})$: 
\begin{equation} 
   D_S\nabla^{2} \boldsymbol{\mu_s} - \frac{\boldsymbol{\mu_s}}{\tau_s} + \omega_{L}\times\boldsymbol{\mu_s}=\mathbf{0} 
\label{eq:Bloch} 
\end{equation} 
which includes spin diffusion: the term with $D_{s}$, spin relaxation: the term with $\tau_{s}$, and spin precession: the term with Larmor frequency $\omega_{L}=\frac{g\mu_{B} B}{\hbar}$, where $g=2$ is the gyromagnetic factor of the spin carrier and $\mu_{B}$ is the electron Bohr magneton.  To solve this equation one sets the boundary condition of an infinite channel: $\mu_{s}\rightarrow 0$ for $d\rightarrow \infty$, where $d$ is the distance from the injector. This is practically fulfilled for $d> 5\lambda_{s}$, due to the fast exponential decay of spin accumulation with $d$.
The spin signal is detected by the ferromagnetic detector which is sensitive to the projection of spin accumulation on its own magnetization direction. It measures the $\mu_{s,y}$ component, which after Ref.~\onlinecite{Fabian2007}, eq.~II.239, reads:
\begin{equation}
\mu_{s,y}(B,d) = \frac{\mu_{s,y}^{0} exp(\frac{-\alpha d}{\lambda_{s}})}{\sqrt{1+(\omega_{L}\tau_{s})^{2}}} (\alpha \cos{\frac{\omega_{L} \tau_{s} d}{2\alpha \lambda_{s}}} - \frac{\omega_{L}\tau_{s}}{2\alpha} \sin{\frac{\omega_{L} \tau_{s} d}{2\alpha \lambda_{s}}})
\label{eq:mu_Hanle}
\end{equation}
where $\alpha=\frac{\sqrt{1+\sqrt{1+(\omega_{L}\tau_{s})^{2}}}}{\sqrt{2}}$ and $\mu_{s,y}^{0} = e P_{i}I \rho \lambda_{s}/(2W)$ is the spin accumulation at the injector at $B=0$, where $e$ is electronic charge, $I$ is the injected current, $P_{i}$ is the polarization of injector, $\rho$ is the resistivity of the channel and $W$ its width. The non-local spin resistance defined as $R_{\text{nl}}=V_{\text{nl}}/I$ is:
\begin{equation}
   R_{\text{nl}}(B, L) = \pm \frac{P_{i} P_{d} \ \rho \ \lambda_{s}}{2W \mu_{s,y}^{0}} \mu_{s,y}(B, L),
   \label{eq:Rnl_Hanle}
\end{equation}
where   $+$ refers to parallel and $-$ to antiparallel alignment of injector/detector magnetization and $P_{d}$ is the polarization of the  detector. $R_{\text{nl}}(B, L)$ from Eq.~\ref{eq:Rnl_Hanle} can be implemented as a fitting function with unknown parameters: $\tau_{s}$, $\lambda_{s}$, $P_{i}P_{d}$, and often one assumes the same polarization for injector and detector $P_{i}=P_{d}=P$. All three parameters can be independently extracted from the fit of the experimental results.

\captionsetup[subfloat]{captionskip=-1.7em,margin=0.1em, justification=raggedright, singlelinecheck=false, font=normalsize, position=top, topadjust=15pt}

\section{\label{sec:analytic_vs_numeric_Hanle}Analytic and numerical model of Hanle effect for confined channel} %
In a device where the channel length is comparable to $\lambda_{s}$ the assumption of vanishing spin relaxation at the device edges is not fulfilled. There are 3 relevant lengths for the problem: spacing between injector and detector $L$, and distances from the injector to the channel ends: $l_{1}$ and $L+l_{2}$, where $l_{\text{dev}} = L+l_{1}+l_{2}$, see Fig.~\ref{fig:spin_random_walk}.
We can view the randomly diffusing spins as being reflected back from the channel ends, see Fig.~\ref{fig:spin_random_walk}. We assume that the reflection from the device edges do not induce extra spin dephasing.   
In this situation the electrochemical potential $\mu_{s}$ at the detector, $d=L$, will interfere, either constructively or destructively, with the  potential created by reflected spins, each effectively traveling larger distances. The resulting signal is the sum over all combinations of trajectories, which pass from $d=0$ to $d=L$, namely ${L , L + 2l_{1}, L + 2l_{2}, L+2l_{1}+2l_{2}, L+2 l_{\text{dev}}},...$. The resulting potential $\Sigma \mu_{s}$ and non-local resistance $\Sigma R_{\text{nl}}$ then reads:
\begin{multline}
   \Sigma \mu_{s}(B,L, l_{1}, l_{2}) = \\
    \sum_{n=0}^{\infty} (\mu_{s}(B,L+2n l_{\text{dev}}) + \mu_{s}(B, L + 2(l_{1} + n l_{\text{dev}}))+ \\
   \mu_{s}(B, L + 2( l_{2}+ n l_{\text{dev}}))+ \mu_{s}(B, L + 2( l_{1}+l_{2}+n l_{\text{dev}}))  
   \label{eq:mu_Hanle_sum}
\end{multline}
and
\begin{equation}
\Sigma R_{\text{nl}}(B, L, l_{1}, l_{2})=\pm \frac{P^{2} \rho \lambda_{s}}{2W \mu_{s,y}^{0}} \Sigma \mu_{s,y}(B,L,l_{1}, l_{2}),
 \label{eq:Rnl_Hanle_sum}
\end{equation}
where the integer $n>0$ counts the passes of spin back and forth through the channel.  As $\mu_{s,y}$ from Eq.~\ref{eq:mu_Hanle} decays exponentially with $d/\lambda_{s}$, one can terminate the sum for trajectories $> 5\lambda_{s}$. The error of such approximation is less than 1\%. One should note that we obtain the same signal when we mirror the device geometry $\Sigma R_{\text{nl}}(B, L, l_{1}, l_{2})=\Sigma R_{\text{nl}}(B, L, l_{2}, l_{1})$. A special case is a semi-infinite device for which only $n=0$ terms in Eq.~\ref{eq:mu_Hanle_sum} are relevant giving  $\Sigma R_{\text{nl}}(B, L, 0,\infty)=2R_{\text{nl}}(B,L)$. The signal is twice bigger than for the channel extending to infinity on both sides.

Alternatively one can define the problem on a discrete grid and solve Eq.~\ref{eq:Bloch} numerically. For that we use a finite element software (COMSOL Multiphysics). We define spin-dependent electrochemical potential as variables $\mu_{\uparrow}$, $\mu_{\downarrow}$ in $x-y$ space, and link them only by spin relaxation. For details see Ref.~\onlinecite{Slachter2011, Wojtaszek2014}. Then, we solve Eq.~\ref{eq:Bloch} at various perpendicular magnetic fields $B$ and various channel geometries $L$, $l_{1}$, $l_{2}$. 
All presented results  are evaluated using typical spin properties for graphene: $D_{s}=0.02$~m$^{2}$/s, $\tau_{s}=200$~ps, $\lambda_{s}=2$~$\mu$m. We set the contact polarization $P=20$~\%, channel resistivity $\rho=150$~$\Omega$ and tunnel barrier resistivity ($\rho_{t}=10$~$\Omega/m$) to avoid the conductivity mismatch problem.

In Fig.~\ref{fig:spin_profile_vsL} we present the numerical evaluation of spin chemical potential $\mu_{s,y}$ across the device for 3 different channel lengths. We can directly see that the presence of the channel ends at distances comparable to $\lambda_{s}$ suppresses the exponential decay and amplifies the magnitude of $\mu_{s,y}$. The numerical model gives exactly the same Hanle precession curves as the standard analytic formula in Eq.~\ref{eq:Rnl_Hanle} only for a very long channel strip where $l_{1}, l_{2} \gg \lambda_{s}$. However, when reducing $l_{1}, l_{2}$, the obtained curves change both in magnitude and lineshape such that they cannot be fitted by Eq.~\ref{eq:Rnl_Hanle}. Instead, there is an excellent agreement in magnitude and in lineshape with the analytic sum of Hanle contributions $\Sigma R_{\text{nl}}(B, L, l_{1}, l_{2})$ from Eq.~\ref{eq:Rnl_Hanle_sum}, where as a cutoff criterion we take $2n l_{\text{dev}}\gtrsim 5\lambda_{s}$. We examined these agreement for various geometries ($l_{1}=l_{2}$, $l_{1}\neq l_{2}$) and for various spin properties: $\lambda_{s}=\{2, 4\ \mu$m$\}$, see Appendix.~\ref{sec:COMSOL_vs_sumHanle}, each time recovering the general scaling of the problem with $\lambda_{s}$.

As the analytic formula is easier to implement all the following results are evaluated using the expression for $\Sigma R_{\text{nl}}$. To reduce the parameter space we vary the channel length evenly on both sides such that $l_{1}=l_{2}=l$.
With the increase of $l_{\text{dev}}$ the amplitude of the spin signal drastically decreases (see Fig.~\ref{fig:Hanle_maximum}) and asymptotically tends to the value for an infinite device. To observe that not only the magnitude of the signal but also its dependence on $B$ changes in Fig.~\ref{fig:norma_Hanle_lineshape} we plot normalized Hanle curves for different $l_{\text{dev}}$ while keeping the same $L$. The differences in Hanle lineshape are most apparent when $L<\lambda_{s}$ and $l_{\text{dev}}\lesssim \lambda_{s}$ as the contribution of reflection from the channel ends increases with confinement. 

\section{\label{sec:Hanle_fit_phase_space} Comparison between standard and extended Hanle formula} 

\captionsetup[subfloat]{captionskip=-1.7em,margin = 0.1em,justification=raggedright,singlelinecheck=false,font=normalsize, position=top, topadjust=15pt}

\begin{figure}[!]
\centering
  \subfloat[]{
   \includegraphics[width=0.32\columnwidth]{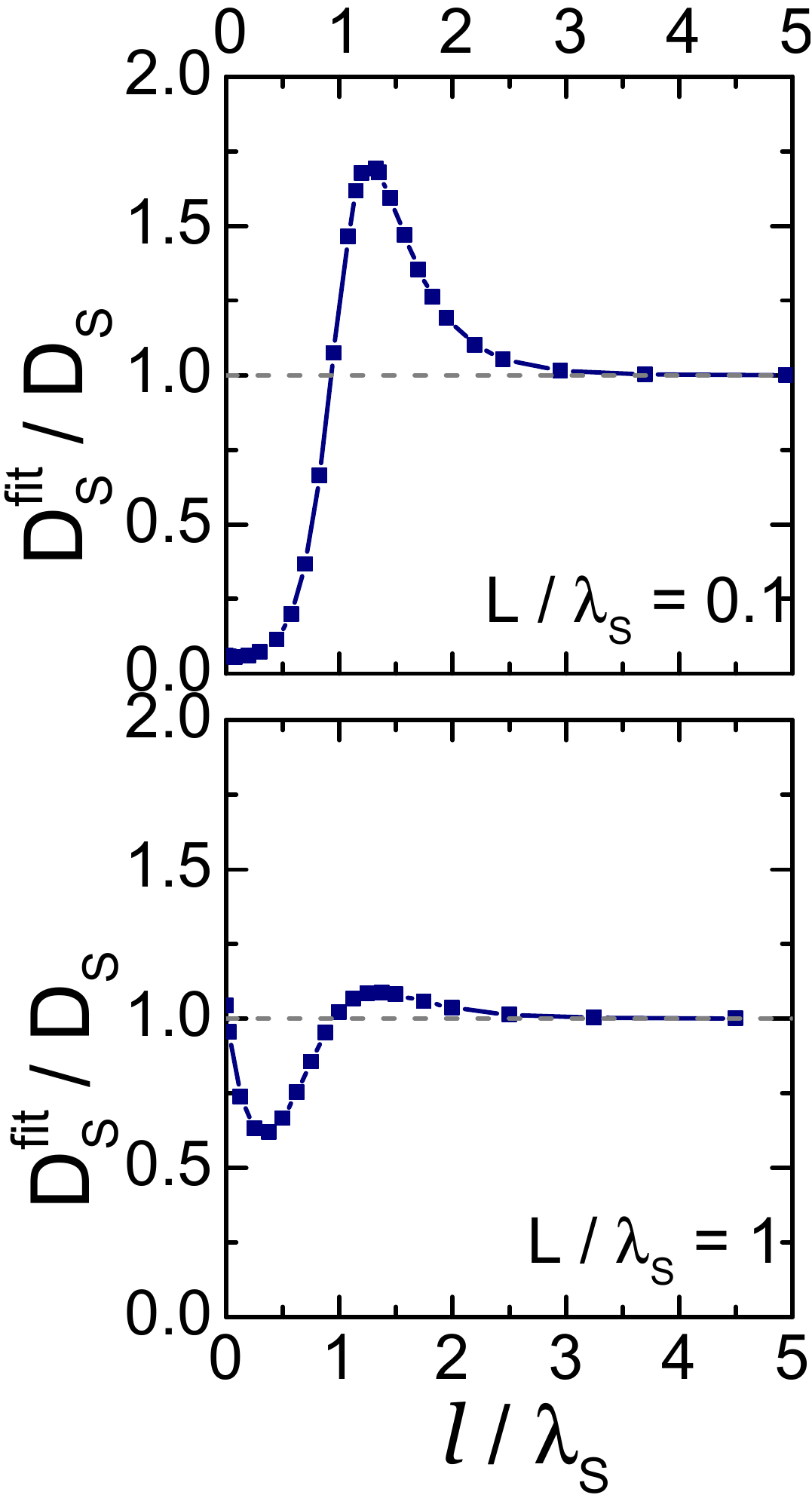} 
   \label{fig:all_fit_Ds}}
    \subfloat[]{
   \includegraphics[width=0.32\columnwidth]{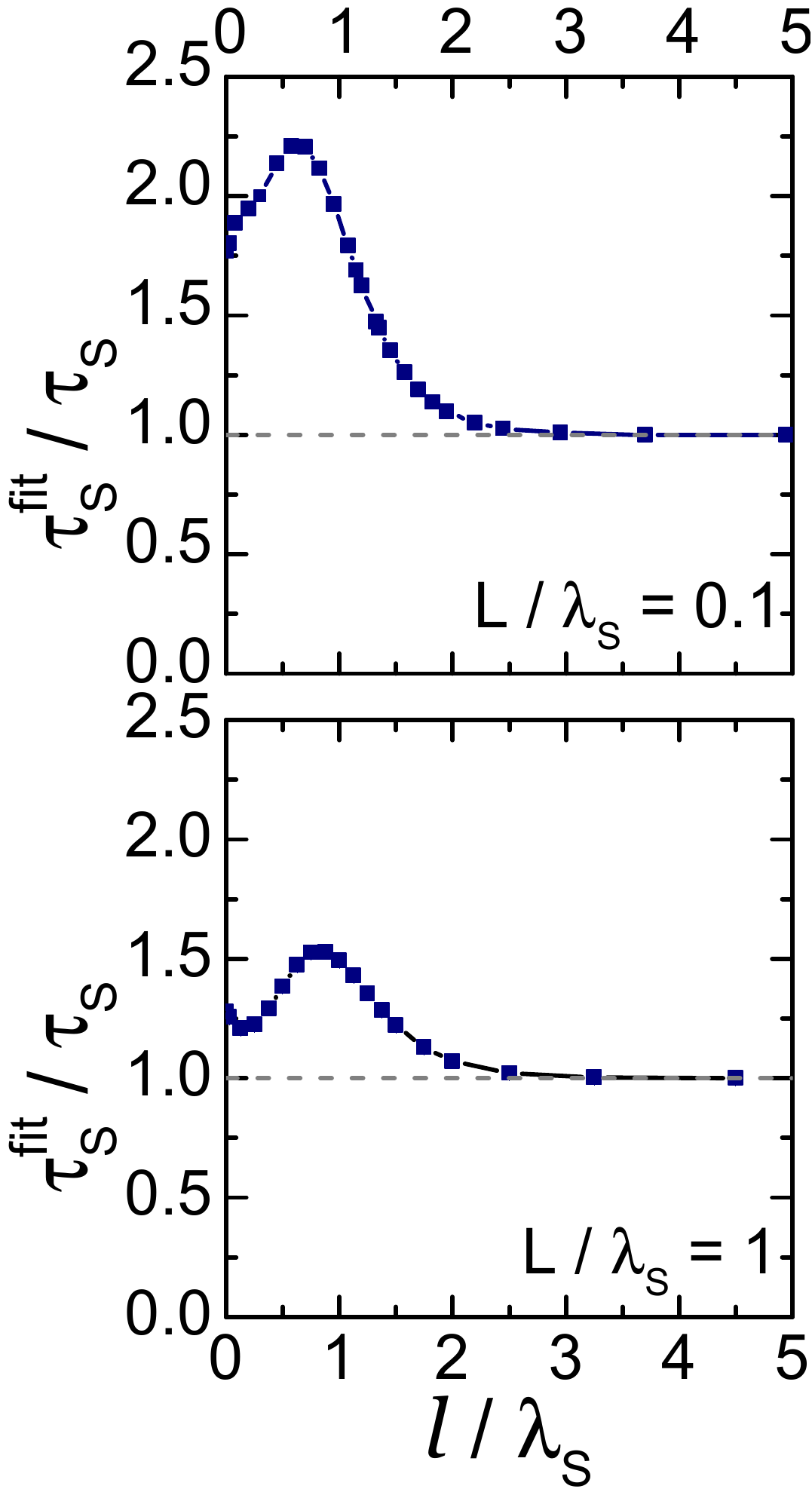} 
   \label{fig:all_fit_tau}}
  \subfloat[]{
   \includegraphics[width=0.32\columnwidth]{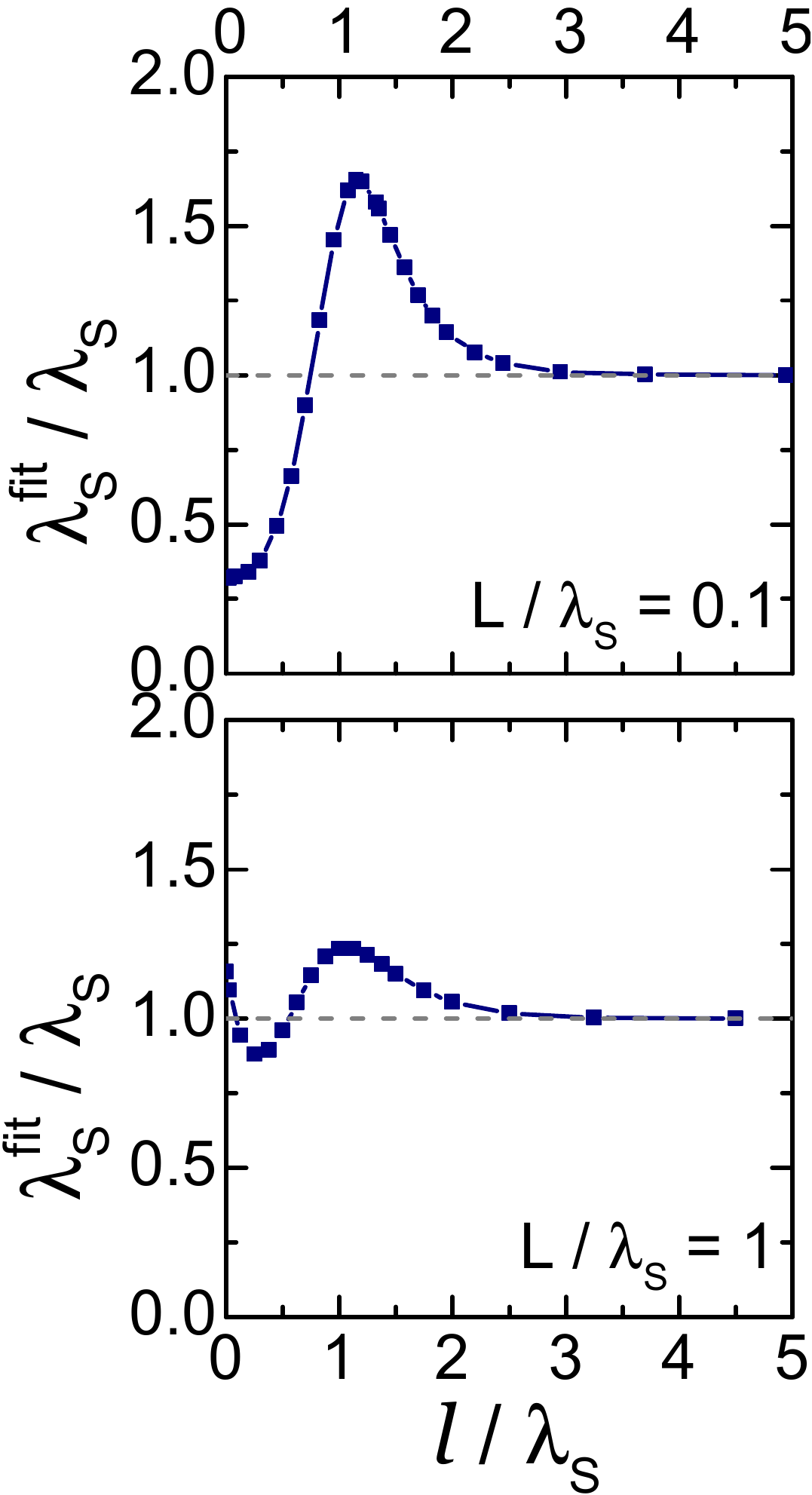} 
   \label{fig:all_fit_lambda}}

\captionsetup{format=hang,justification=centerlast, font=small}
\caption[]{(Color online) (a) $D_{s}^{\text{fit}}$ (b) $\tau_{s}^{\text{fit}}$ and (c) $\lambda_{s}^{\text{fit}}$ extracted from fitting of $\Sigma R_{\text{nl}}(B,L,l,l)$ with  $R_{\text{nl}}(B,L)$ from Eq.~\ref{eq:Rnl_Hanle} for two different cases: $L=0.1 \lambda_{s}$ (upper) and $L=1 \lambda_{s}$ (lower). The fitted coefficients are normalized by the true spin properties of the channel. 
}
\label{fig:Hanle_all_fit}
\end{figure} 
Next, we analyze the discrepancy between true and extracted spin coefficients when enforcing a standard Hanle fit $R_{\text{nl}}$ using Eq.~\ref{eq:Rnl_Hanle} to the signal from a device of finite length properly described by the sum $\Sigma R_{\text{nl}}$ from Eq.~\ref{eq:Rnl_Hanle_sum}. The accuracy of the fit depends on $L$ and $l$ and their relation to $\lambda_{s}$. We first 
construct Hanle lineshapes of the form $\Sigma R_{\text{nl}}(B, L, l, l)$ for different geometries $(L,l)$ and individually fit them using the $R_{\text{nl}}$ from Eq.\ref{eq:Rnl_Hanle}. The fitting uses the least squares method for magnetic field range of (-0.4, 0.4)T and the extracted coefficients $D_{s}^{\text{fit}}$, $\tau_{s}^{\text{fit}}$ and $\lambda_{s}^{\text{fit}}$ are collected irrespective of the visual discrepancy between the $\Sigma R_{\text{nl}}$ and the fitting curve. In Fig.~\ref{fig:Hanle_all_fit} we present the ratio between true and extracted spin properties for two cases:  $L=0.1 \lambda_{s}$ and $L=1 \lambda_{s}$. For large~$l$ the spin coefficients $D_{s}^{\text{fit}}$, $\tau_{s}^{\text{fit}}$ and $\lambda_{s}^{\text{fit}}$ converge to the true spin properties of the channel $D_{s},\ \tau_{s}, \ \lambda_{s}$. However for $l<3\lambda_{s}$ there is a strong modulation of the extracted coefficients due to the interference mechanism between spins multireflected from the channel ends. The maximum discrepancy in $\tau_{s}^{\text{fit}}$, irrespective of the $L/\lambda_{s}$ ratio, appears for $l\simeq\lambda_{s}$. In this range and for $L=0.1 \lambda_{s}$, $\tau_{s}^{\text{fit}}$ gets overestimated by more than a factor of 2, but the corresponding discrepancy for $D_{s}^{\text{fit}}$ is about 70\% smaller. Increasing the ratio of $L/\lambda_{s}$ improves the accuracy of the fit, because the contribution from the reflected spins vanishes.  $D_{s}^{\text{fit}}$ tends to zero when $L,l \ll \lambda_{s}$, which is the case for a strongly confined structure. Further we show that in this range the spin transport is zero-dimensional, with a characteristic Lorentzian dependence.

\section{\label{sec:Lorentz_fit_0D}The limit of 0D transport}
When $l_{\text{dev}}\ll \lambda_{s}$, the spin chemical potential distributes evenly along the channel due to the spin backflow/multireflection and reduces the process of diffusion. Without diffusion one enters a 0-dimensional transport regime, a case more familiar in optical spin polarization experiments in semiconductors \cite{Lou2006} and three-terminal (3T) Hanle precession measurements \cite{Dash2009}. The 0D steady state behavior is described by a linear equation with only relaxation and precession term \cite{Dyakonov2008}:
\begin{equation}
   \frac{\boldsymbol{\mu_{s}}}{\tau_{s}}=\omega_{L} \times \boldsymbol{\mu_{s}}
\end{equation}   
and its solution is the Lorentzian function: $\mu_{s,y} = \frac{\mu_{s,y}^{0}}{1+(\omega_{L}\tau_{s})^2}$.  In Appendix~\ref{sec:Lorentz_from_Hanle} we verify the correlation between the Lorentzian dependence and $\Sigma R_{\text{nl}}$ from Eq.~\ref{eq:Rnl_Hanle_sum} for a strongly  confined channel.  
  
\begin{figure}
\centering
  \subfloat[]{  
   \includegraphics[width=0.45\columnwidth]{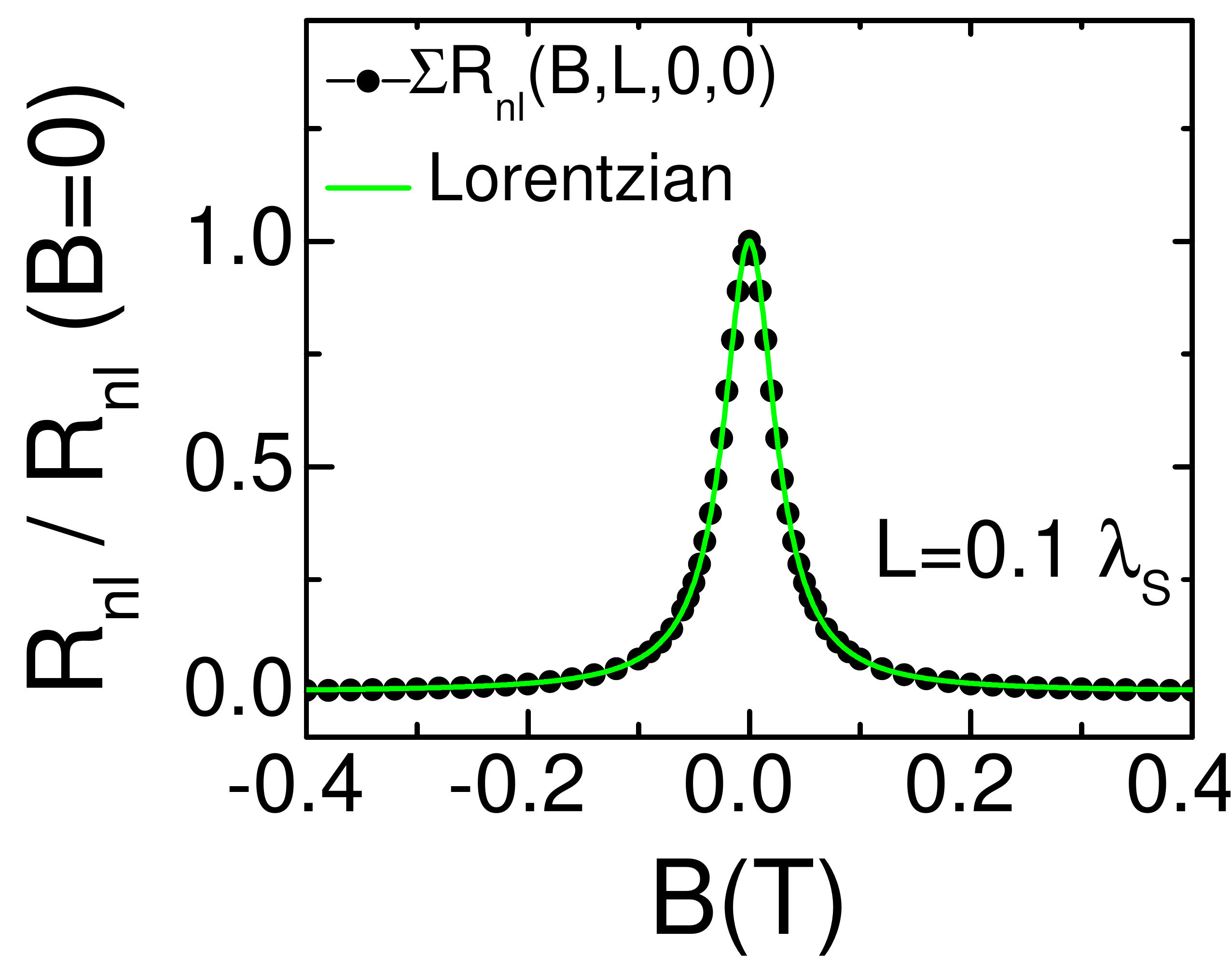} 
   \label{fig:Lorentzian_fit}}
     \quad 
  \subfloat[]{  
   \includegraphics[width=0.45\columnwidth]{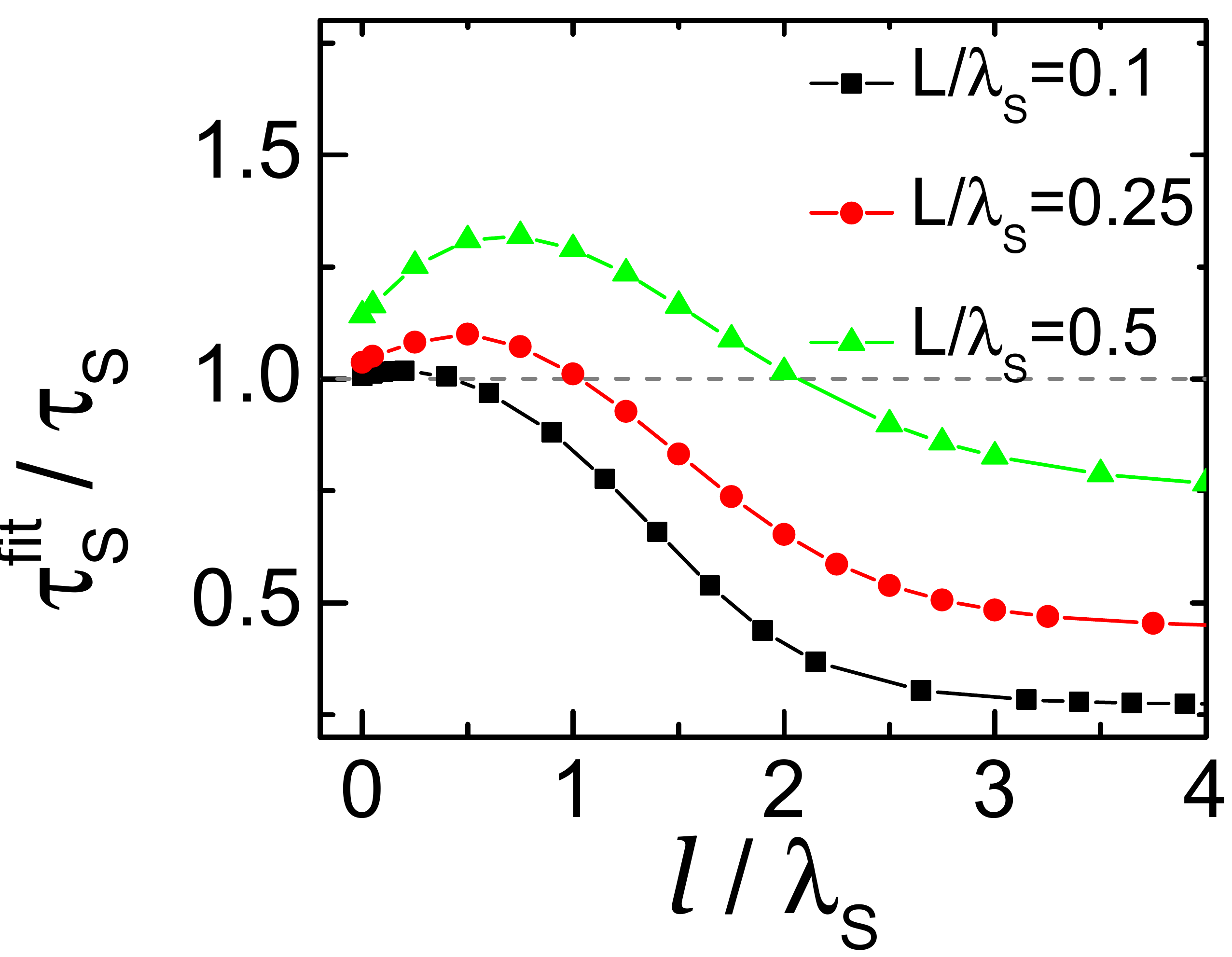} 
   \label{fig:Lorentzian_vs_x}}
     \qquad 
\captionsetup{format=hang,justification=centerlast, font=small}
\caption[]{(Color online) (a) Sum of Hanle signals $\Sigma R_{\text{nl}}(B, 0.1\lambda_{s},0,0)$ for the confined spin channel and its least square fit to the Lorentzian curves. For the Lorentzian fit $\tau_{s}^{\text{fit}}\simeq \tau_{s}$, whereas $\tau_{s}^{\text{fit}}\simeq 2\tau_{s}$ when standard Hanle fit from Eq.~\ref{eq:Rnl_Hanle} is used. (b) The  $\tau_{s}^{\text{fit}}$ extracted from Lorentzian fit of $\Sigma R_{\text{nl}}$ for $L<\lambda_{s}$, normalized by the true  $\tau_{s}$ of the channel. 
}
\label{fig:Lorentz_all_fit}
\end{figure}

In Fig.~\ref{fig:Lorentzian_fit} we present a comparison between the Hanle signal for confined channel $\Sigma R_{\text{nl}}(B, 0.1\lambda_{s}, 0,0)$ and its Lorentzian fit.  There is a very good agreement in lineshape between these two curves although their  amplitudes are different, see Appendix \ref{sec:Lorentz_from_Hanle}. In experimental situations, however, only the lineshape, which determines the $\tau_{s}$, is relevant. 
Next we perform the same Lorentzian fitting for different $\Sigma R_{\text{nl}}(B, L, l,l)$, where $L<\lambda_{s}$, to establish the geometrical range where the quasi-Lorentzian dependence still holds and gives a good estimation of $\tau_{s}$, see Fig.~\ref{fig:Lorentzian_vs_x}. Indeed for $l_{\text{dev}}<\lambda_{s}$ and $L\lesssim\lambda_{s}/4$ there is a good agreement between the true $\tau_{s}$ and $\tau_{s}^{\text{fit}}$. Therefore the transport in that range can be considered as zero-dimensional.   

\captionsetup[subfloat]{captionskip=-1.7em,margin = 0.1em,justification=raggedright,singlelinecheck=false,font=normalsize, position=top, topadjust=15pt}

Note that the 0D transport description does not apply to the 3T Hanle measurements in devices with long spin channel and narrow injecting contact. In such a case diffusion from the injector plays a role and Eq.~\ref{eq:mu_Hanle} with $d=0$ should be used in the fitting procedure instead of the Lorentzian function. The 3T resistance then reads:
\begin{equation}
  R_{\text{3T}}(B)=\pm \frac{P_{i}P_{d}\rho \lambda_{s}}{2\sqrt{2}W} \frac{\sqrt{1+\sqrt{1+(\omega_{L} \tau_{s})^{2}}}}{\sqrt{1+(\omega_{L}\tau_{s})^{2}}}.
\end{equation}

\section{Interpretation of the spin transport measurements}
The finite length of the spin channel largely affects the magnitude of the spin accumulation, Hanle precession lineshape and the extracted fitting coefficients. 
From the above analysis we find that in the case of tunneling contacts, when injector/detector are located far from the channel ends, or $l_{1}, l_{2}>3\lambda_{s}$, we can apply (even for $L\lesssim0.1 \lambda_{s}$) the standard Hanle formula from Eq.~\ref{eq:Rnl_Hanle} for 1D spin transport system. For the case when $L<0.25 \lambda_{s}$ and $l_{\text{dev}}\lesssim \lambda_{s}$ we can use a zero-dimensional description with Lorentzian formula and extract $\tau_{s}$ with  less than 15\% discrepancy. For all the ranges in between the error in extracted coefficients largely depends on specific lengths of $L$, $l_{1}$, $l_{2}$ in relation to $\lambda_{s}$ and one needs to use $\Sigma R_{nl}$ from Eq.~\ref{eq:Rnl_Hanle_sum} as a fitting function. 
In practice the spin coefficients extracted from standard Hanle fitting  using Eq.~\ref{eq:Rnl_Hanle} can be applied for a crude approximation of the true spin coefficients, and when the device geometry in relation to extracted $\lambda_{s}$ requires, one should repeat the fit with few additional Hanle terms in Eq.~\ref{eq:Rnl_Hanle_sum}, keeping the same $D_{s}$, $\tau_{s}$, $P$ for all of them. 

It is important to note that using the presented Hanle lineshape analysis one can also investigate how the reflection from the edges and channel ends affects the spin interference pattern and determine the validity of the assumption about non-dephasing edges \cite{Popinciuc2009, Idzuchi2012, Yazyev2008}.
 
\section{\label{sec:Conclusions}Conclusions} 
We discuss how a finite length of the spin channel leads to a spin confinement and spin interference, affecting two characteristics of the spin transport, namely spin accumulation and Hanle precession lineshape. To properly account for these geometrical effects we extend the standard Hanle formula, which depends only on the distance between injector/detector $L$, into a sum of Hanle functions, which depends also on distances to the channel ends $l_{1}$, $l_{2}$.  We confirm this model numerically by solving the spin Bloch equation on a finite device geometry.
Further we analyze the discrepancy one gets when fitting the signal from a confined channel by the standard Hanle formula derived for infinite channel length. The error in extracted $\tau_{s}^{\text{fit}}$ can be more than 2-fold when the device length is comparable to $\lambda_{s}$ and $L\lesssim 0.1 \lambda_{s}$. In the regime of very short spin channel the spin diffusion is strongly reduced and zero-dimensional spin transport dominates. We present how the proposed sum of Hanle curves reflects this transition and 
show the conditions when a fit to the zero-dimensional Lorentzian lineshape yields an accurate spin lifetime.  

This work provides an useful insight into the interference effects in finite-size devices in Hanle precession and into the transition between 1D and 0D spin transport.

\medskip
\begin{acknowledgments} 
We would like to acknowledge A.~Slachter and F.~L.~Bakker for introducing to us COMSOL modeling. This work was financed by NanoNed, the Zernike Institute for Advanced Materials, the Netherlands
Organisation for Scientific Research (NWO), the Foundation for Fundamental Research on Matter (FOM) and the European Union 7$^{th}$
Framework Programme under grant agreement n$^{\circ}$604391 Graphene Flagship. 
\end{acknowledgments}

\appendix
\section{\label{sec:COMSOL_vs_sumHanle} Analytic versus numerical Hanle signal in finite device.}
\begin{figure}[!h]
\centering
  \subfloat[]{  
   \includegraphics[width=0.44\columnwidth]{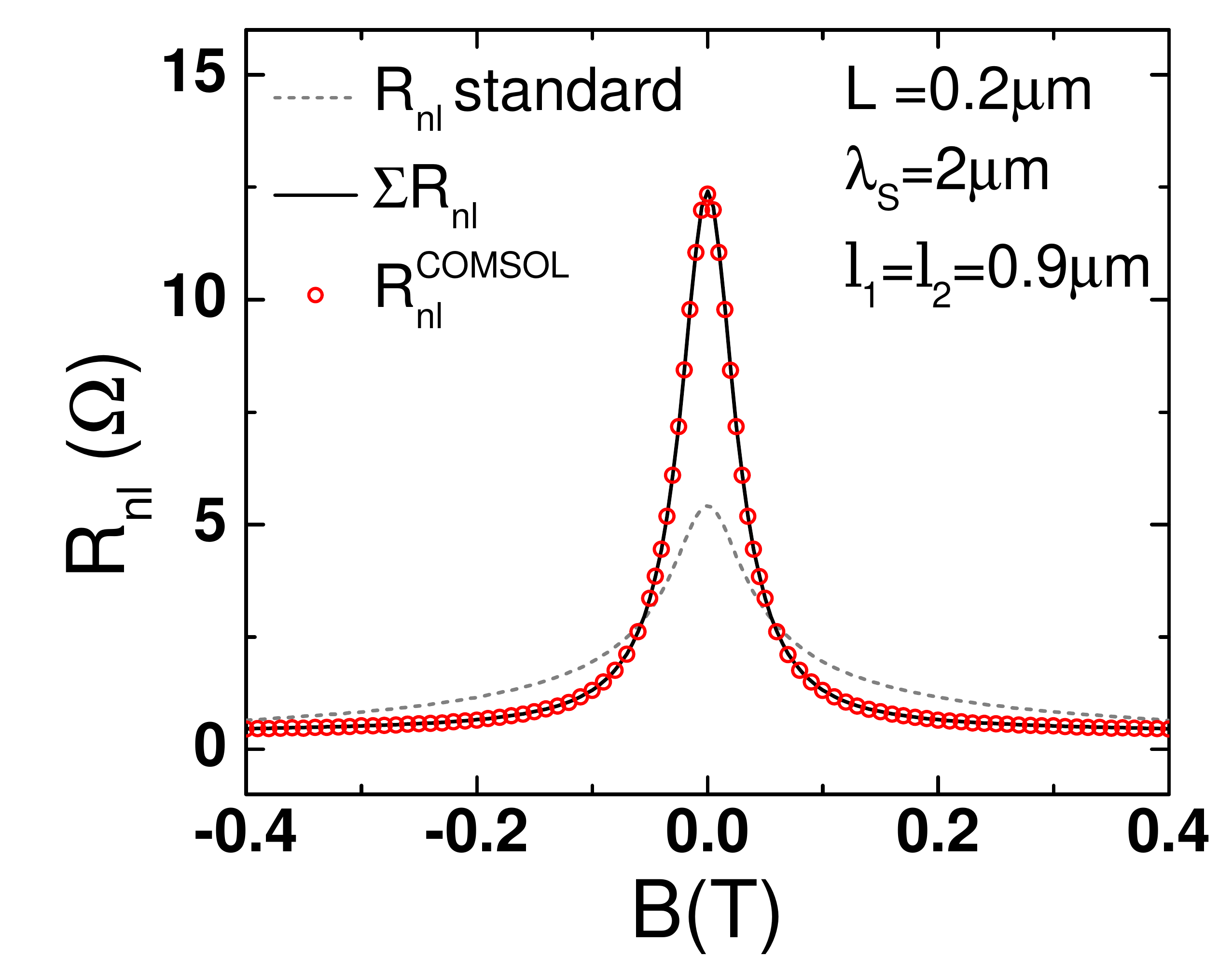} 
   \label{fig:comp_Hanle_sym}}
     \quad 
  \subfloat[]{  
   \includegraphics[width=0.44\columnwidth]{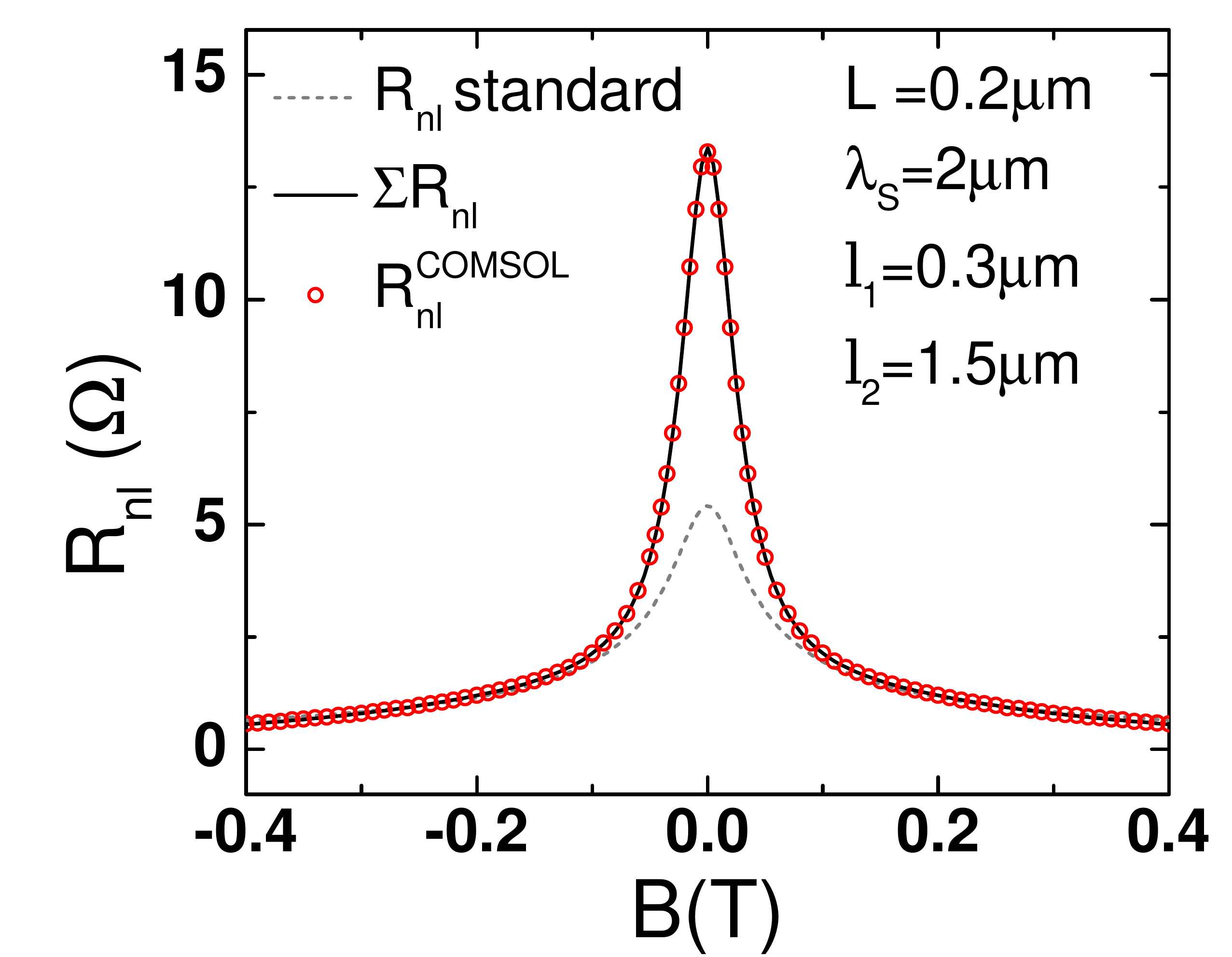} 
   \label{fig:comp_Hanle_asym}}
     \qquad 
\captionsetup{format=hang,justification=centerlast, font=small}
\caption[]{(Color online) Comparison between numerical and two analytic expressions for Hanle effect: $R_{nl}(B,L)$ from Eq.~3 and $\Sigma R_{\text{nl}}(B, L,l_{1},l_{2})$ from Eq.~5 in main text. (a) Case when the injector-detector are spaced symmetrically from the channel ends $l_{1}=l_{2}$. (b) Case when the injector-detector are located asymmetrically from the channel ends $l_{1}\neq l_{2}$. The spin properties of the channel are: $D_{s}=0.02$~m$^{2}$/s, $\tau_{s}=200$~ps, $\lambda_{s}=2\ \mu$m, $P_{i}=P_{d}=0.2$, $\rho=150$~$\Omega$.
}
\end{figure}

Here we compare the Hanle signal obtained from finite-element simulation of spin transport in COMSOL\cite{Wojtaszek2014}, where the device geometry is set explicitly, with the introduced analytic summation formula $\Sigma R_{\text{nl}}(B, L, l_{1}, l_{2})$ after Eq.~5 in main text. We truncate the sum in Eq.~5 using the criterion $d>5\lambda_{s}$ as in the main text. The channel electronic and spin properties in both numerical and analytic cases are set the same and the non-invasive nature of contacts in COMSOL is assured by placing highly resistive tunneling barrier between contact and the channel. We examine the agreement for various geometries: $l_{1}=l_{2}$, $l_{1}\neq l_{2}$, and for various spin properties: $\lambda_{s}=\{2, 4\ \mu$m$\}$ and always obtain an excellent matching both in the magnitude and in the lineshape of Hanle signal. In Fig.~\ref{fig:comp_Hanle_sym} and \ref{fig:comp_Hanle_asym} we present two examples of such an agreement for the cases of symmetrically and asymmetrically spaced contacts from the channel ends, where $L=0.1 \lambda_{s}$ and graphene length $l_{\text{dev}}=l_{1}+L+l_{2}$ is the same.  For reference we also plot the standard Hanle signal for infinite channel length, $R_{\text{nl}}(B, L)$ after Eq.~3, which, what is also shown in the main text, does not agree neither in amplitude nor in lineshape to the signal for a confined channel.

\captionsetup[subfloat]{captionskip=-1.7em,margin = 0.1em,justification=raggedright,singlelinecheck=false,font=normalsize, position=top, topadjust=15pt}

\section{\label{sec:Lorentz_from_Hanle} Transition from Hanle to Lorentzian dependence in strongly confined systems}
In 0D case there is no diffusion of spins and spin dynamics reduces to a linear equation with only relaxation and precession term of which solution is the Lorentzian function: $\mu_{s,x} = \frac{\mu_{s,x}^{0}}{1+(\omega_{L}\tau_{s})^2}$. 
The very same Lorentzian dependence is obtained for the three-terminal (3T) spin transport measurements \cite{Dash2009}, when the size of the injecting contact is much larger than $\lambda_{s}$ and all reflection effects from the device ends can be ignored. In 3T the injecting contact is at the same time a voltage probe so that the measured signal can be viewed as sum of standard Hanle signals from Eq.~2 for $L$ equal all possible distances between injection and detection points within the contact. The continuous nature of the contact makes $L$ also a continuous variable and the sum of Hanle signals forms an integral $\int_{0}^{\infty} \mu_{s,y}(B,x)dx$. Using the fact that $\int_{0}^{\infty} \text{exp}(-ax)\cos{(bx)} dx=a/(a^{2}+b^{2})$ and $\int_{0}^{\infty} \text{exp}(-ax)\sin{(bx)} dx=b/(a^{2}+b^{2})$ one can show that:
\begin{align}
\sum_{n=0}^{\infty} \mu_{s, y}&(B,n L)  \xrightarrow{L/\lambda_{s}\rightarrow 0} \nonumber \\
& \int_{0}^{\infty} \mu_{s,y}(B,x\lambda_{s})dx =\frac{\mu_{s,x}^{0}}{1+(\omega_{L}\tau_{s})^2}.
\label{eq:HtoL_convergence}
\end{align}
 
Unlike for the 3T case, we consider in our work a device which has narrow contacts, but in addition the channel length is also very small, $l_{\text{dev}}<\lambda_{s}$ leading to strong spin confinement. The expression for $\Sigma \mu_{s,y}(B, L, l_{1}, l_{2})$ from Eq.~4 is very similar to the discretized version of Eq.~\ref{eq:HtoL_convergence}, when $L, l_{1}, l_{2} \ll \lambda_{s}$. However, there are some subtle differences. For example in the case of $\Sigma \mu_{s,y}(B,L,0,0)=4\mu_{s,y}(B, L)+4\mu_{s,y}(B, 3L)+\mu_{s,y}(B, 5L)+...$, the sum contains only contributions from odd terms of $L$  while the Lorentzian sum from Eq.~\ref{eq:HtoL_convergence} contains both odd and even terms $L$, e.g. $\mu_{s,y}(B, L)+\mu_{s,y}(B, 2L)+\mu_{s,y}(B, 3L)+\mu_{s,y}(B, 4L)+\mu_{s,y}(B, 5L)...$. In numerical integration we can approximate these missing terms by the preceding terms, $\mu_{s,y}(B, (2n+1)L)\approx \mu_{s,y}(B, 2nL)$, so the sum $\Sigma \mu_{s,y}(B,L,0,0)$ will be twice bigger than the corresponding Lorentzian sum, Eq.~\ref{eq:HtoL_convergence}, but the characteristic lineshape will be preserved. When the distances $l_{1}, l_{2}>0$ then the
the correlation of the sum  $\Sigma \mu_{s,y}(B, L, l_{1}, l_{2})$  with a Lorentzian is less straightforward, but for $l_{1}, l_{2}<\lambda_{s}$ the terms for ${L , L + 2l_{1}, L + 2l_{2}, L+2l_{1}+2l_{2}, L+2 l_{\text{dev}}},...$ can be seen as a change in numerical quadrature and will also give a quasi-Lorentzian dependence. 


\end{document}